\documentclass[twocolumn,showkeys,preprintnumbers,amsmath,amssymb]{revtex4}
\usepackage{graphicx}

\usepackage{fancyhdr}

\begin{document}


\title{ESTIMATION OF THE AMOUNT OF THE NUCLEAR TRANSFORMATION PRODUCTS
FORMED UNDER EXPLOSION-INDUCED COMPRESSION OF A SUBSTANCE TO THE
SUPERDENSE STATE}

\author{S. V. Adamenko}
\author{A. S. Adamenko}
\author{I. A. Kossko}
\author{V. D. Kurochkin}
\altaffiliation{Institute of Problems of Materials Science of the
NAS of Ukraine, 3 Krzhizhanovsky Str., Kyiv, 03142 Ukraine}
\author{V. V. Kovylyaev}
\altaffiliation{Institute of Problems of Materials Science of the
NAS of Ukraine, 3 Krzhizhanovsky Str., Kyiv, 03142
Ukraine}
\author{S. S. Ponomarev} \altaffiliation{Institute of
Problems of Materials Science of the NAS of Ukraine, 3
Krzhizhanovsky Str., Kyiv, 03142 Ukraine}
\author{A. V. Andreev}
\altaffiliation{Shevchenko Kyiv National University, 64
Volodymyrs'ka Str., Kyiv, 01033 Ukraine}

\affiliation{%
Electrodynamics Laboratory ``Proton--21''\\
14/1 Dovjenko Str., 03057, Kiyv, Ukraine}%
\email{enr30@enran.com.ua}

\date{\today}

\begin{abstract}%
In the present work carried out at the Electrodynamics Laboratory
``Proton--21'' with the use of X-ray electron probe microanalysis
(XEPMA) and glow-discharge mass spectrometry (GDMS) techniques, we
study the chemical composition of a substance formed as a result
of the explosion-induced compression of solid targets and
deposited on the surfaces of accumulating screens. We established
that the explosion products contain chemical elements which were
not included in the composition of the initial materials of
targets and accumulating screens or were included in them as
impurities in quantities by 3\dots{}7 and more orders in magnitude
less than those detected after the experiment. We show that the
appearance of ``new'' chemical elements on the surfaces of
accumulating screens is not connected, first, with their
redistribution from the bulk of a very accumulating screen, a
target, or structural details of the experimental chamber which
participated in the explosion initiation and, secondly, is not
caused by the processes of deposition from the residual atmosphere
of the vacuum chamber or by the transfer from the shell walls
(from the structural details which were present in the
experimental chamber but did not participate in the process of
explosion). The estimated values of the total mass and the total
number of atoms of all the chemical elements, which appeared as
the result of the explosion of one target manufactured of pure Cu,
or Pb, or Ag and were located only on the surface of the most
enriched central part of an accumulating screen of about 5\,mm in
diameter, are approximately $1 \times 10^{-4}$\,g and $1.2 \times
10^{18}\dots 1.7\times 10^{18}$~atoms, respectively. We conclude
that the regular appearance of chemical elements, which were not
included in the composition of the initial materials of targets
and accumulating screens, in the explosion products is the
consequence of a nuclear transformation of a part of their
material, i.e., is the fact testifying to both the running of the
nucleosynthesis reactions upon the explosion-induced destruction
of targets and to the first realized possibility of controlled
creation of the conditions for their running in a laboratory
setup.
\end{abstract}


\maketitle  \thispagestyle{fancy}

\section*{INTRODUCTION}
  From 1999 till the present time, the staff of the Electrodynamics
Laboratory ``Proton--21'' carries on the experiments on the
explosion-induced compression of a substance, which leads to the
creation of superdense states. Till October 2003, over 5000
dynamical impact compressions of solid targets were performed at
the Laboratory. The goal of these experiments is the verification
of the preliminarily theoretically justified hypothesis of the
possibility to initiate, by using the pulse coherent action on a
solid substance, the self-organizing process of avalanche-like
self-densification up to the state of collapse (of the
electron-nucleus plasma), in which the conditions for the running
of collective many-particle nuclear reactions arise due to the
effective screening of the Coulomb barrier. For the verification
of this hypothesis, the Laboratory's staff constructed the
experimental setup which is able to transfer up to 1\,kJ of energy
to a solid target for a pulse duration of about 10\,ns with the
help of the electron beam used as a primary carrier of the
concentrated energy. At the culmination stage of the process, the
microvolume of the target substance was compressed up to a density
of above $10^{26}$\,cm$^{-3}$. In this case, the power density in
the region of compression exceeded, by various estimations,
$10^{22}$\,W/cm$^{3}$.

\begin{figure*}
\centering
\includegraphics[width=16 cm]{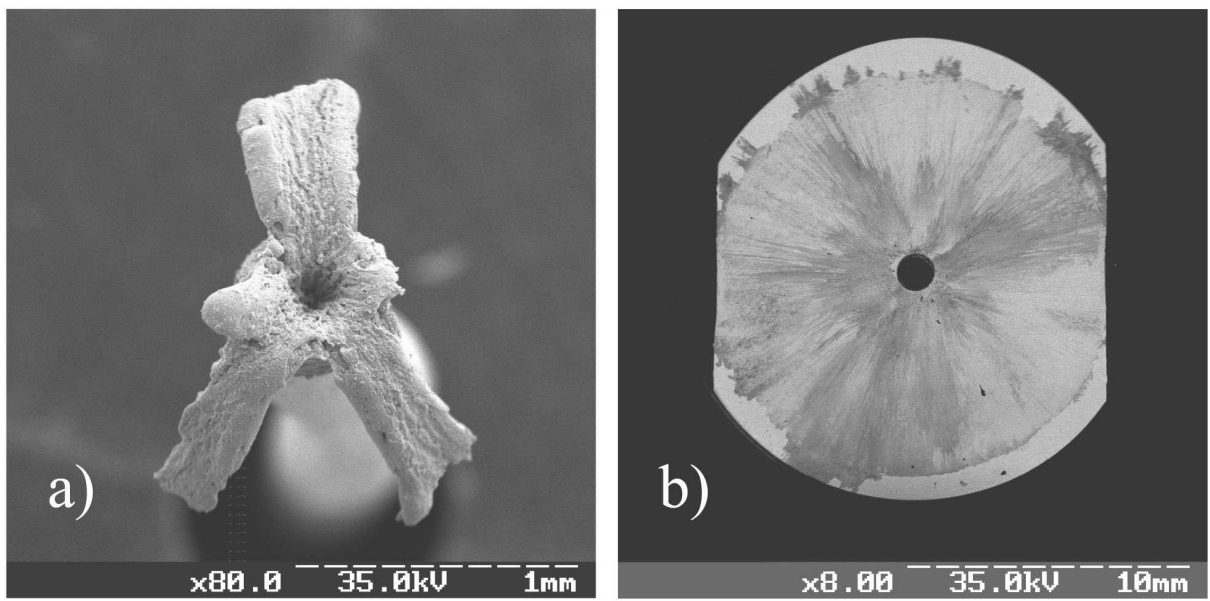}
\caption{Scanning electron images (the secondary-electron mode) of
a typical target after the explosion (a) and a typical
accumulating screen with deposited products of the explosion (b).}
\label{Fig_est01}
\end{figure*}

  The impact compression of a target was realized in vacuum of
about $10^{-3}$\,Pa and led to its fracture by the explosion from
inside. This process was usually accompanied by the radial
dispersion of a target material with its deposition on a special
accumulating disk-like screen of about 15\,mm in diameter and
0.5\,mm in thickness. Fig.~\ref{Fig_est01} presents the photos of
both a typical target after the explosion (a) and a typical
accumulating screen with deposited products of the explosion (b)
which were derived with a scanning electron microscope in the
secondary-electron mode. By using the method of scanning electron
microscopy (SEM), we established that the products, which remained
in the target crater and precipitated on accumulating screens,
formed a layer of irregularly distributed drops, splashes, films,
particles, and other micro- and nanoobjects with complicated
morphology.

  Because the explosion products are microobjects, we firstly studied
them mainly with local methods. To investigate their element and
isotope compositions, we used X-ray electron probe microanalysis
(XEPMA), local Auger electron spectroscopy (AES), laser mass
spectrometry (LMS), and secondary-ion mass spectrometry (SIMS). By
using the methods of mass spectrometry, we discovered deviations
of the isotope composition from the natural abundance of isotopes
for a number of chemical elements contained in these
products~\cite{Mes_Adam,Visnyk}. While studying the chemical
composition of products of the explosion by XEPMA, AES, and SIMS,
we registered up to several tens of chemical elements in
significant amounts in every analysis [including those cases where
both a target and an accumulating screen were made of a single
chemical element maximally purified from impurities, e.g., Cu
(99.99 mass. \%), Ag (99.99 mass. \%), Pb (99.75 mass. \%), and
others]. These elements were not found by high-sensitivity
analytical methods in the initial materials of targets and
accumulating screens or were present in them as impurities but at
concentrations by 3\dots 7 orders and more less than those
measured after the experiment~\cite{Mes_Adam,Visnyk}. Moreover,
while studying the specimens by AES, we registered a collection of
Auger-peaks which cannot be referred, on the one hand, to
artefacts of the analysis and, on the other one, be related to any
Auger-peaks of the known chemical elements~\cite{Mes_Adam}. The
mentioned Auger-peaks were referred by us in the scope of the
known part of the Periodic table to basically unidentifiable
peaks. As one of the variants of the interpretation of the
revealed unidentifiable Auger-peaks, the assumption as for their
affiliation to long-lived transuranium elements was discussed
in~\cite{Mes_Adam}.

\begin{figure}
\centering
\includegraphics[width=8 cm]{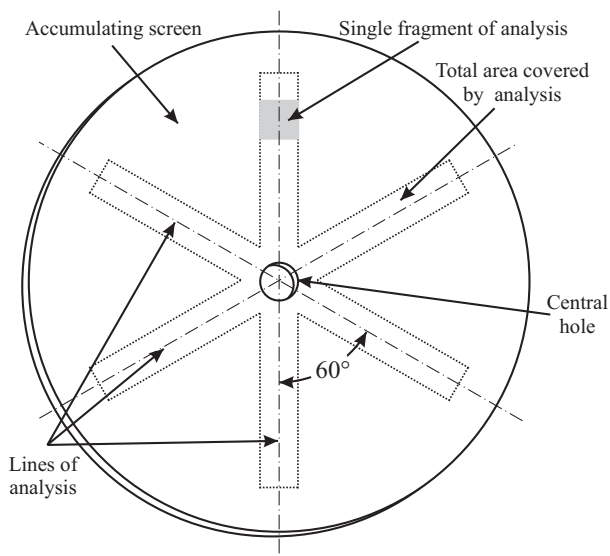}
\caption{Scheme of the estimation of the total number of particles
on the whole accumulating screen No. 5094.} \label{Fig_est02}
\end{figure}

  In authors' opinion, all the above-presented facts point unambiguously to
the running of the artificially initiated intense processes of
nucleosynthesis and transmutation of chemical elements in the
microvolume of the target substance undergone to the impact
compression to superhigh densities. The key fact is, undoubtedly,
the appearance of chemical elements, which were absent in the
composition of the initial materials of targets and accumulating
screens (the structural details of the experimental chamber which
participated in the process of explosion), in the explosion
products. The justification of this fact is the main goal of the
present work.

\section{MATERIALS AND METHODS}

  To solve the posed problem, we used such methods of quantitative
determination of the element composition of a substance as XEPMA
and GDMS.

  The studies based on XEPMA were carried out on an X-ray REMMA--102
microanalyzer (SELMI, Sumy, Ukraine) equipped with two wavelength
dispersion X-ray spectrometers and one energy-dispersion X-ray
spectrometer [with a Si(Li) detector]. Spectra were registered at
a 35-keV accelerating voltage of the electron beam, a probe
current of 0.1 nA, and a residual pressure of $2 \times
10^{-4}$\,Pa in the specimen chamber. The range of energies
registered by an energy-dispersion spectrometer was 0.9\dots
30\,keV, the energy resolution on the line MnK$_\alpha$ at the
counting rate up to 1000\,pulse/sec was 150\,eV, and the typical
time of the registration of spectra was 200\dots 400\,sec. For the
quantitative analysis, we used a standard computer program for the
calculation of the concentrations of elements developed by the
firm-producer of the device (SELMI).

  To analyze the element composition, we used a high-sensitivity
glow-discharge mass spectrometer (with Ar plasma) VG~9000 (VG
Elemental, UK). The current and voltage of discharge were,
respectively, 1.8\,mA and 1.1\,kV. As a holder, we used a cell for
plane specimens without cooling which ensures the analyzed region
diameter to be 5\,mm. The residual pressures in the specimen
chamber and in the spectrometer were, respectively, at most
$1\times 10^{-2}$ and $1 \times 10^{-5}$\,Pa. The ion beam was
accelerated by a voltage of 8\,kV. The spectrometer possesses the
range of analyzed masses 1\dots 250, and its mass resolution
$M/\Delta M$ at the half-height of the Cu mass-peak is at the
level of 7000\dots 9000. For the quantitative analysis, we used a
standard computer program for the calculation of the
concentrations of elements developed by the firm-producer of the
device (VG~Elemental).

  In the present work, we studied the accumulating Cu screens in
both the initial state and after the deposition of the explosion
products on them (see Fig.~\ref{Fig_est01}, b). In fact, a screen
served as a substrate. The explosion products were deposited on
one of its surfaces as a layer which possessed a weakly pronounced
relief and was characterized by the axial symmetry. In the central
part of an accumulating screen, we can see an area in the form of
a pit of $10\dots 20\,\mu$m in depth relative to the screen
surface and about $5\dots 7$\,mm in diameter
(Fig.~\ref{Fig_est01}, b
, and \ref{Fig_est06}). Its formation was caused by the
entrainment of the screen material as a result of the explosion of
a target. Accumulating screens were used to be as-received, i.e.,
they were not undergone, prior to the analysis, to any damaging
cleaning procedures or those changing their composition. As
materials for targets in the present study (wires of 0.5\,mm in
diameter; see Fig.~\ref{Fig_est01}, a), we chose Cu, Pb, and Ag.

  For the solution of special tasks, we manufactured assemblies
or ``sandwiches'' gathered from 20--30 accumulating screens
closely fitting one another. In this case, we carried out the
analysis on the lateral side of an assembly, i.e., on the edges of
accumulating screens. Such a structure of specimens and the
analysis scheme guarantee a sufficient averaging over
inhomogeneities of the composition of materials of the very
accumulating screens and over those of the explosion products. We
note that a homogeneous specimen was necessary for the correct
application of such a destructive method of analysis as GDMS, in
which the registration of the ion currents of various chemical
elements (isotopes) was implemented by the successive scheme.

\section{RESULTS AND DISCUSSION}

  The studies of the composition of products of the explosion of
targets, the determination of chemical elements appeared due to
the explosion, and the estimation of the number of atoms of such
chemical elements were carried out at the Laboratory beginning
from the first stages of the Project. The objects of these studies
were the layer formed by the explosion products deposited on an
accumulating screen or the layer remained in the target crater
after the explosion. At first, we believe that local methods of
analysis are most suitable for studying these objects. Indeed, the
chemical elements which are contained in trace amounts in
microobjects can be easily detected by local methods and, at the
same time, be beyond the detection limit of even very sensitive
integral methods. However, further, the necessity to use also
integral methods of analysis becomes obvious. By using namely
integral methods for the comparison of the composition of a whole
accumulating screen prior to and after the explosion of a target
and by considering the composition of the target material
transferred on the accumulating screen, we can prove that the
appearance of chemical elements in the explosion products which
were absent in perceptible quantities in the initial materials of
both the target and accumulating screen is the result of
nucleosynthesis rather than that of their redistribution from the
bulk of the accumulating screen or target.

  At the same time, we note that even if the use of integral methods
solves the posed problem on the whole, their results do not
include completely the results derived by local methods. On the
one hand, local methods give the important information as for the
abundance area of newly appeared chemical elements and the
character of their distribution there. On the other one, the
comparison of the estimations of a required quantity which were
derived by different methods supplementing one another seems to be
very interesting and useful.

  On the XEPMA-based determination of the number of atoms of the chemical elements
appeared on an accumulating screen as a result of the target
explosion, we took pure copper (Cu, 99.99 mass \%) as the material
of both the target and accumulating screen. As an example, we
consider the study of the surface layer of accumulating screen No.
5094. The estimation of the number of atoms of the foreign
chemical elements (except for Cu) was realized in two stages. At
first, we counted the number of atoms in the particles lying on
the screen surface and then in the enriched $2\dots 3\,\mu$m-thick
surface layer of the matrix; the values derived in this way then
were summed. The indicated thickness of the enriched surface layer
of the matrix was evaluated from the data on the profiles of
concentrations through depth of accumulating screens by SIMS.

  Below we describe the procedure and results of the first stage.
The scheme is presented in Fig.~\ref{Fig_est02}. An analyzed area
was a raster (a square) of $54.3\times54.3\,\mu$m in size. On it,
we counted the number of all particles, analyzed their composition
with an sharp probe, and determined the number of atoms for each
foreign chemical element at each particle. Then the analyzed area
was shifted by a step equal to its side length along one of the
lines where the analysis was performed. Such lines formed an angle
of $60^{\circ}$ with one another (see Fig.~\ref{Fig_est02}).
Having performed this procedure, we registered 417 spectra derived
on different particles. By virtue of the axial symmetry, we
assumed that all the lines of analysis are equivalent and total
analyzed area along them is representative for the whole
accumulating screen.

  We estimated the total number of particles on the whole accumulating
screen surface as
\begin{equation}
N_{p.s.} = \frac{S_s}{S_a}N_{p.a.} \approx 2.0 \times 10^5,%
\label{Eq01}
\end{equation}
where $S_s$ and $S_a$ are, respectively, the total area of the
accumulating screen and the total area of the analyzed region, and
$N_{p.a.}=417$ is the total number of the analyzed particles. We
determined the numbers $N_{ij}$ of atoms of the $i$-th foreign
chemical element in the $j$-th particle. Then we calculated the
mean number of atoms of the $i$-th foreign chemical element per
particle by the formula
\begin{equation}
\overline{N}_{i} = \frac{ \sum\limits_{j=1}^{N_{p.a.}} {N_{ij} } }{ N_{p.a.}}.%
\label{Eq02}
\end{equation}

With regard for Eq.~\ref{Eq02}, we were able to estimate the total
number of atoms of the \textit{i}-th foreign chemical element
contained in the particles on the whole accumulating screen
surface as
\begin{equation}
{N}_{i} = \overline{N}_{i}\times { N_{p.s.}},%
\label{Eq03}
\end{equation}

The values derived as the result of processing the spectra are
given in Table~\ref{Table1}.

\begin{table}
\centering%
\caption{\label{Table1}Numbers of atoms of the foreign chemical
elements contained in all particles placed on the whole surface of
accumulating screen No. 5094.}
\begin{tabular}{cccc}
\hline \hline%
        &Number of&        &Number of\\[-0.9ex]
Element &atoms per &Element &atoms per \\[-0.9ex]
        &specimen &        &specimen\\
\hline%
Mg      &3.06E+15&Y       &2.04E+14\\
Al      &9.08E+16&Zr      &2.75E+13\\
Si      &3.19E+16&Ag      &6.14E+15\\
P       &9.07E+15&Cd      &2.20E+15\\
S       &1.94E+16&In      &1.92E+15\\
Cl      &6.70E+16&Sn      &1.61E+16\\
K       &2.19E+16&Te      &1.39E+15\\
Ca      &1.28E+16&Ba      &2.43E+15\\
Ti      &3.48E+15&La      &7.16E+14\\
V       &5.08E+13&Ce      &2.51E+15\\
Cr      &2.40E+15&Pr      &1.52E+14\\
Mn      &5.89E+14&Ta      &4.15E+15\\
Fe      &5.11E+16&W       &2.27E+16\\
Co      &3.88E+14&Au      &5.67E+15\\
Ni      &2.07E+14&Pb      &1.90E+17\\
Zn      &2.87E+16&        &        \\
\hline
        &        &\textbf{TOTAL}   &5.99E+17\\

\hline \hline
\end{tabular}
\end{table}

  Finally, by summing all values given in Table~\ref{Table1}, we got the total
number of atoms of the foreign chemical elements contained in all
particles placed on the whole surface of accumulating screen No.
5094 as
\begin{equation}
N_{\Sigma{}p.} = \sum\limits_{i}{N_{i}} \approx 5.99\times 10^{17}.%
\label{Eq04}
\end{equation}

  By an similar scheme, we determined the number of atoms of
the foreign chemical elements contained in the enriched surface
layer of the accumulating screen matrix. In this case, we chose a
raster to be $11\times11\,\mu$m in size. The less area was taken
in order to more easily find the areas of the screen surface
positioned along the lines of analysis and containing no
particles. We registered the spectra over the whole area of a
raster and the number of these spectra registered from different
analyzed areas was 113. We recall that the elementary analyzed
region of the surface layer of the accumulating screen was
$11\times11\,\mu$m in size with a thickness of about $3\,\mu$m
(the range of depths of the specimen from which we registered
X-ray emission).

  As above, first, we calculated the amount of the foreign
chemical elements in the analyzed region and then, with regard for
the ratio of the area of the the screen surface to that of the
analyzed region, recounted in the amount of the foreign chemical
elements contained in the surface layer of the matrix of the whole
screen. The results are given in Table~\ref{Table2}.

\begin{table}
\centering%
\caption{\label{Table2}Numbers of atoms of the foreign chemical
elements contained in the surface layer of the matrix of the whole
accumulating screen No. 5094.}
\begin{tabular}{cccc}
\hline \hline%
        &Number of&        &Number of\\[-0.9ex]
Element &atoms on &Element &atoms on \\[-0.9ex]
        &specimen &        &specimen\\
\hline%
Al      &2.13E+17&Ca      &5.44E+15\\
Si      &6.62E+16&Mn      &8.77E+14\\
P       &1.09E+16&Fe      &3.57E+16\\
S       &1.01E+17&Zn      &4.63E+16\\
Cl      &7.60E+16&Pb      &6.22E+15\\
K       &3.10E+16&        &        \\
\hline%
        &        &\textbf{TOTAL}  &5.93E+17\\
\hline \hline
\end{tabular}
\end{table}

  As above, by summing the values from Table~\ref{Table2}, we get that the
total number of atoms of the foreign chemical elements contained
in the surface layer of the matrix of the whole accumulating
screen is
\begin{equation}
N_{\Sigma{}m.} \approx 5.93\times 10^{17}.%
\label{Eq05}
\end{equation}

Now with regard for Eq.~\ref{Eq04} and Eq.~\ref{Eq05}, we get that
the total number of atoms of the foreign chemical elements
appeared on accumulating screen No. 5094 as the result of the
explosion of a target and the number of nucleons in them are,
respectively,
\begin{eqnarray}
N_\Sigma & = & N_{\Sigma{}p.}+N_{\Sigma{}m.} \approx 1.2\times
10^{18}, \label{Eq06}\\
N_\mathrm{nucl.} & = & 8.33 \times 10^{19}. \label{Eq07}
\end{eqnarray}

  Analyzing the results derived (see Eq.~\ref{Eq04} and Eq.~\ref{Eq05}), we note that
the character of the distribution of foreign chemical elements
appeared on the accumulating screen surface is such that exactly a
half of them is contained in the particles placed on it and the
other half belongs to the surface layer of the matrix of at most
$3\,\mu$m in thickness. It is easy to calculate that the mentioned
number of foreign atoms corresponds to their concentration in the
analyzed surface layer of the accumulating screen to be about 3
mass \%. With regard for the purity of the initial materials of
the accumulating screen and target (99.99 mass \%), the last fact
means that it is senseless to correct the derived value for the
content of impurities in the initial material of the screen and in
the transferred material of the target.

  Thus, the derived value does correspond to the number of foreign
atoms appeared on the accumulating screen as the result of the
explosion of a target. But if we conclude, by basing on this value
that the appeared foreign atoms were generated only in the course
of nucleosynthesis, such an assertion can be hardly named
sufficiently strict. In our opinion, two quite weak points are
present in the above reasoning. On the one hand, the derived value
is not the result of a direct measurement, but it is based on a
number of statistical hypotheses and model ideas of the morphology
of the surface layer of the accumulating screen whose degree of
reliability and adequacy can be, generally saying, called in
question. In other words, it is very difficult to estimate the
accuracy of the derived value, though we are sure that its order
is estimated properly. On the other hand, the very procedure of
derivation of the required value does not exclude theoretically
the possibility of that the foreign atoms can appear on the
surface layer of the accumulating screen through the
redistribution of impurities from the bulk of the very
accumulating screen as the result of target explosion rather than
be the result of a nuclear transformation of its material.

  In our second investigation, we tried to overcome the mentioned
drawbacks of the first experiment. For the analysis of the element
composition, we used the highly sensitive method of GDMS. By its
gist, this method is integral (we chose the diameter of an
analyzed region to be 5\,mm), and its application with the purpose
to find the number of foreign atoms appeared on the accumulating
screen does not require to use any model ideas of the morphology
and structure of its surface. In other words, in the evaluation of
the required quantity, the method can be used in such a way that
this quantity will be found as the result of a direct measurement.

  As for the effect of redistribution of the composition of an
accumulating screen, we can take it into account, for example, if
we take the scheme of analysis such that, within it, the
composition of the whole accumulating screen is registered rather
than that of the surface layer. In this case, the composition of
the accumulating screen should be registered twice: in the initial
state and then after the explosion of a target. If the mentioned
compositions will be identical, the enrichment of the surface
layer of the screen occurs due to the redistribution of its
composition over the specimen volume. But if the content of minor
elements in the accumulating screen composition increases after
the explosion of a target, we can say about the appearance of the
atoms of foreign chemical elements and count their amount.

\begin{figure}
\centering
\includegraphics[width=8 cm]{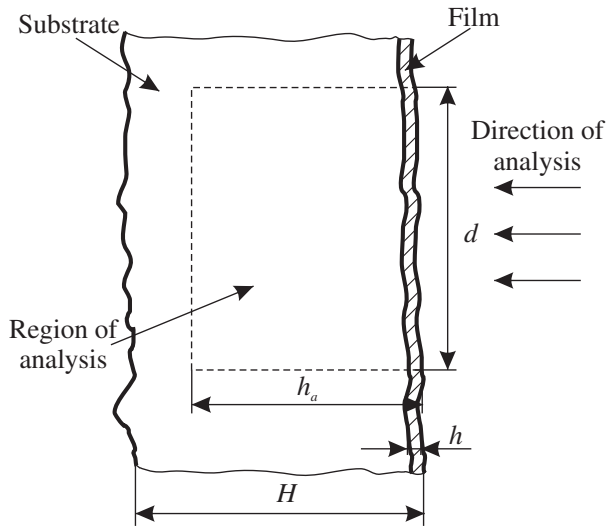}
\caption{Scheme of the cross section of an accumulating screen.}
\label{Fig_est03}
\end{figure}

  Consider the proposed scheme of analysis. Fig.~\ref{Fig_est03} shows schematically
the cross section of an accumulating screen. There we also drew
the region of analysis in the case where the analysis procedure
begins from that side of the screen on which the film from
products of the explosion of a target is deposited. It is obvious
that the depth $h_a$ of the analyzed region varies depending on
the duration of etching. In the case where the condition
\begin{equation}
h_a \leq h,%
\label{Eq08}
\end{equation}
holds, where $h$ is the film thickness, the results of measurement
correspond to the film composition. If $h_a$ satisfies the
condition
\begin{equation}
h < h_a < H,%
\label{Eq09}
\end{equation}
where $H$ is the specimen thickness, the film composition is
affected by that of the substrate in the increasing degree, and
the results of measurement lose any significant physical sense.
Finally, consider the situation with
\begin{equation}
h_a = H.%
\label{Eq10}
\end{equation}

In this case, the results of measurement correspond to the
composition of an accumulating screen. The indicated case can be
uniquely characterized with the geometric factor $k_0$ equal to
the ratio of the area of a cross section of the film $S_f$ to the
area of a cross section of the whole accumulating screen $S_0$. It
is obvious that $k_0$ satisfies the relation
\begin{equation}
k_0=\frac{S_f}{S_0}=\frac{d\cdot h}{d\cdot H}=\frac{h}{H}.%
\label{Eq11}
\end{equation}

  In view of the real situation, we may take the values of 2 and
$500\,\mu$m for the film thickness $h$ and the specimen thickness
$H$, respectively. Then the geometric factor
\begin{equation}
k_0=0.004.%
\label{Eq12}
\end{equation}

  At first sight, it seems that we can determine the total composition
of the accumulating screen following the presented scheme of
analysis if relation (Eq.~\ref{Eq10}) is satisfied. However, its
realization meets some difficulties related to the fact that a
mass spectrometer with magnetic mass analyzer is constructed so
that only those ions can be registered at a given moment which are
characterized by a specific nominal mass number defined by the
magnetic induction of the field in the mass analyzer. Therefore,
in order to register the whole mass spectrum of a specimen, we
need to successively scan the whole mass range. In this case, of
course, it is necessary that the specimen under study be
homogeneous by composition, at least, through depth in order that
the registered spectrum have any physical meaning. Otherwise, we
can find themselves in the situation, e.g., where we etch a
specimen throughout and fail to register any mass-peak in the
spectrum. Such a situation would occur in the case of a layered
specimen, in which the layers from pure chemical elements are
placed in the order of a decrease in their mass numbers with
increase in depth, i.e., in the order inverse to that of scanning
the mass range by the magnetic analyzer.

\begin{figure}
\centering
\includegraphics[width=8 cm]{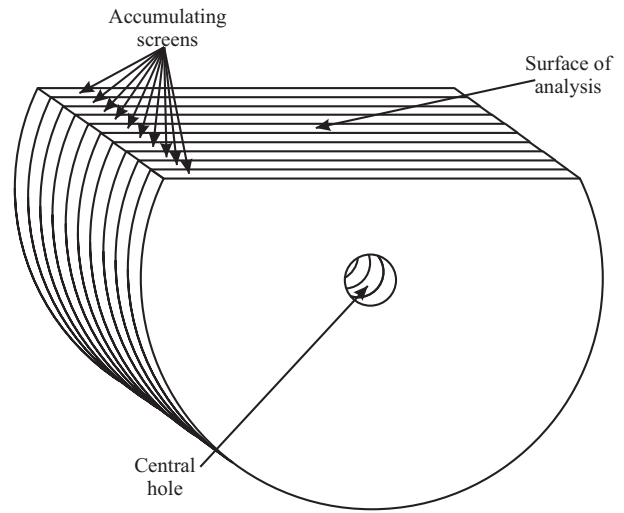}
\caption{Scheme of a ``sandwich'' type assembly constructed from
accumulating screens.} \label{Fig_est04}
\end{figure}

  The above-indicated difficulties can be avoided in the following
way. We prepared the specimen as an assembly made of several
accumulating screens. It was constructed so that, first, it had a
homogeneous composition in the direction from the analyzed surface
through depth and, secondly, the analyzed region on the analyzed
surface possessed the geometric factor $k_0$. If both these
conditions are satisfied, the procedure of registration of
mass-spectra is correct and the results of analysis reflect the
composition of accumulating screens.

  Fig.~\ref{Fig_est04} represents schematically a variant of an assembly made
of accumulating screens: the so-called ``sandwich'' which
obviously satisfies the first condition mentioned above. Indeed,
while passing through depth from the analyzed surface even at a
distance of the order of several hundreds of $\mu$m, this specimen
can be considered to be homogeneous with quite high accuracy.

\begin{figure}
\centering
\includegraphics[width=8 cm]{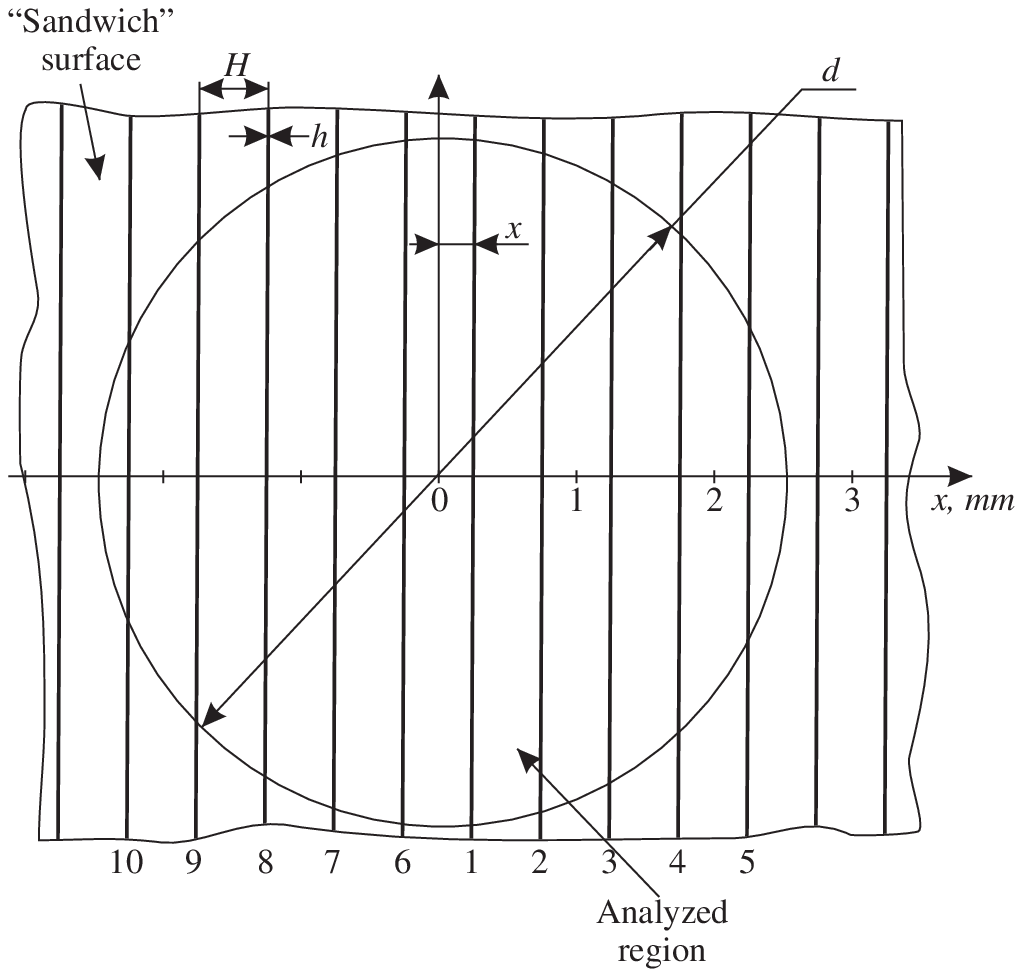}
\caption{Scheme of analysis on a specimen of the ``sandwich''
type, where $H$ is the accumulating screen thickness, $h$ is the
thickness of a film formed by products of the explosion of a
target, $d$ is the diameter of the analyzed region, $x$ is a
displacement of the film relative to the center of the analyzed
region, 1--10 is the numbers of films in the analyzed region ($H =
500\,\mu$m, $h = 2\,\mu$m, and $d = 5000\,\mu$m).}
\label{Fig_est05}
\end{figure}

  At last, we should like to clarify the situation concerning the
geometric factor upon the determination of the composition of a
specimen of the ``sandwich'' type. The analysis scheme is given in
Fig~\ref{Fig_est05}. It is obvious that the geometric factor for a
``sandwich'' satisfies the relation
\begin{equation}
k_s=\frac{S_{fs}}{S_a},%
\label{Eq13}
\end{equation}
where $S_{fs}$ is the cross section area of all films being in the
analyzed region and $S_a$ is the area of the analyzed region.
Since $d = 5000\,\mu$m, we get
\begin{equation}
S_a=\frac{\pi d^2}{4}=19\,634\,954\,\mu\mathrm{m}^2.%
\label{Eq14}
\end{equation}

It is obvious that
\begin{equation}
S_{fs}=h L,%
\label{Eq15}
\end{equation}
where $L$ is the length of all film layers being in the analyzed
region. Now the entire problem is reduced to the determination of
$L$. In Fig.~\ref{Eq05}, all the layers being in the analyzed
region are enumerated from 1 to 10. Assume that the 1st layer is
shifted relative to the center of the analyzed region by $x$.
Then, we can write
\begin{equation}
L(x)=\sum\limits_{i=1}^{10}{l_i(x)},%
\label{Eq16}
\end{equation}
where $l_i(x)$ is the length of the $i$-th layer, which can be
found by the Pythagoras theorem. By substituting Eq.~\ref{Eq16} in
Eq.~\ref{Eq15}, we can find $S_{fs}$. It turns out that $S_{fs}$
does not depend on $x$ and has the same value for any $x$ from the
interval $0<x<H$:
\begin{equation}
S_{fs}=78540\,\mu\mathrm{m}^2.%
\label{Eq17}
\end{equation}

Finally, by substituting Eq.~\ref{Eq17} and Eq.~\ref{Eq14} in
Eq.~\ref{Eq13}, we find the value of the geometric factor $k_s$ on
a ``sandwich'':
\begin{equation}
k_{s}=0.004.%
\label{Eq18}
\end{equation}

By comparing Eq.~\ref{Eq18} and Eq.~\ref{Eq12}, we see that
\begin{equation}
k_{s}=k_0.%
\label{Eq19}
\end{equation}

Thus, we may infer that a construction of the ``sandwich'' type
specimen and the proposed scheme of analysis satisfy two
conditions formulated above, which means that we can correctly
determine the composition of accumulating screen in the
``sandwich'' type specimen.

\begin{table}
\centering%
\caption{\label{Table3}Data on mass losses in the experiments with
Pb targets and Cu screens.}
\begin{tabular}{ccc}
\hline \hline%
Specimen &\multicolumn{2}{c}{Mass loss, mg}        \\
\cline{2-3}
 number  &Screen  &Target \\
\hline%
6383 & 4.73 & 4.86\\
6799 & 3.22 & 1.72\\
6775 & 4.90 & 2.16\\
6776 & 5.11 & 1.44\\
6778 & 5.09 & 3.52\\
6779 & 4.39 & 3.79\\
6783 & 3.85 & 1.09\\
6784 & 3.04 & 3.13\\
6788 & 4.41 & 3.02\\
6793 & 3.75 & 1.24\\
\hline%
Average  &4.25    &2.60    \\
\hline \hline
\end{tabular}
\end{table}

  In the second investigation, the initial material of accumulating
screens was, as earlier, Cu, and we took Pb for targets. By the
GDMS data, the total content of impurities in lead $C_{im.t.}$ was
0.04 mass \%. By the results of weighting, the mean losses of the
target mass after the explosion, $\Delta m_t$, and the mass of the
accumulating screen dispersed by the explosion plasma, $\Delta
m_s$, were about 2.60 and 4.25~mg, respectively (see
Table~\ref{Table3}). The ``sandwich'' was gathered from 20 halves
of accumulating screens processed by explosions (see Table 3),
whose blanks were cut from the same Cu sheet. As the specimen of
the initial material of an accumulating screen, we took blank No.
6815 cut from the same sheet.

  The composition of the initial material of an accumulating screen
was determined four times by GDMS on the blank mentioned above.
The averaged results of measurements are presented in
Table~\ref{Table4}, where we also show the composition of the
accumulating screen after the explosion of the Pb target which was
averaged over the results of 4 measurements on the ``sandwich.''
All 4 measurements were performed on the same place in the central
region of the analyzed surface on the ``sandwich'' (see
Fig.~\ref{Fig_est04}) successively one after another, i.e., the
determined composition corresponds to the central area of the
accumulating screen.

\begin{table*}
\centering%
\caption{\label{Table4}Change of the element composition of a
accumulating Cu screen (``sandwich'') after the explosion of a Pb
target.}
\begin{tabular}{cccccccc}
\hline \hline%
Element &\multicolumn{3}{c}{Concentration, mass \%} &Element& \multicolumn{3}{c}{Concentration, mass \%}       \\
\cline{2-4} \cline{6-8}
        &Initial  &``Sandwich''& Increment& &Initial  &``Sandwich''& Increment \\
\hline%
H & 1.30E-06 & 4.81E-06 & 3.52E-06 & Rh & 1.26E-05 & 6.43E-05 & 5.17E-05\\
He & 3.81E-06 & 4.22E-06 & 4.11E-07 & Pd & 1.97E-05 & 4.44E-05 & 2.47E-05\\
Li & 2.63E-06 & 1.33E-05 & 1.07E-05 & Ag & 2.48E-03 & 2.90E-03 & 4.17E-04\\
Be & 1.48E-06 & 2.19E-04 & 2.18E-04 & Cd & 7.38E-05 & 1.31E-04 & 5.68E-05\\
B & 2.64E-05 & 1.16E-04 & 8.95E-05 & In & 6.83E-06 & 1.21E-05 & 5.26E-06\\
C & 3.80E-04 & 2.22E-03 & 1.84E-03 & Sn & 6.53E-04 & 2.26E-04 & -4.27E-04\\
N & 5.80E-04 & 4.55E-03 & 3.97E-03 & Sb & 4.68E-04 & 3.86E-04 & -8.15E-05\\
F & 7.28E-05 & 1.83E-04 & 1.10E-04 & I & 1.66E-06 & 1.27E-05 & 1.11E-05\\
Ne & 1.67E-06 & 2.25E-06 & 5.84E-07 & Te & 1.48E-04 & 2.65E-04 & 1.17E-04\\
Na & 1.98E-04 & 7.21E-03 & 7.01E-03 & Xe & 1.48E-05 & 5.63E-05 & 4.15E-05\\
Mg & 2.49E-05 & 2.60E-04 & 2.35E-04 & Cs & 2.46E-06 & 3.50E-06 & 1.03E-06\\
Al & 1.99E-06 & 2.48E-04 & 2.46E-04 & Ba & 6.94E-05 & 2.12E-05 & -4.82E-05\\
Si & 7.50E-07 & 4.57E-04 & 4.56E-04 & La & 1.23E-06 & 1.64E-06 & 4.04E-07\\
P & 1.80E-02 & 2.40E-02 & 5.96E-03 & Ce & 1.39E-06 & 2.80E-06 & 1.40E-06\\
S & 3.71E-03 & 5.38E-03 & 1.67E-03 & Pr & 1.33E-06 & 4.72E-06 & 3.39E-06\\
Cl & 1.08E-03 & 6.54E-02 & 6.43E-02 & Nd & 8.70E-06 & 1.42E-05 & 5.51E-06\\
K & 6.71E-07 & 1.15E-04 & 1.15E-04 & Sm & 8.60E-06 & 1.69E-05 & 8.31E-06\\
Ca & 6.91E-04 & 4.25E-03 & 3.56E-03 & Eu & 2.57E-06 & 4.41E-06 & 1.84E-06\\
Sc & 1.17E-05 & 8.49E-05 & 7.32E-05 & Gd & 8.63E-06 & 1.05E-05 & 1.84E-06\\
Ti & 8.11E-06 & 1.23E-03 & 1.23E-03 & Tb & 1.16E-06 & 2.28E-06 & 1.12E-06\\
V & 1.24E-06 & 4.39E-05 & 4.27E-05 & Dy & 6.60E-06 & 6.06E-06 & -5.48E-07\\
Cr & 1.97E-05 & 9.22E-06 & -1.05E-05 & Ho & 1.32E-06 & 2.53E-06 & 1.21E-06\\
Mn & 4.77E-05 & 8.67E-05 & 3.91E-05 & Er & 5.46E-06 & 5.25E-06 & -2.09E-07\\
Fe & 4.76E-03 & 1.15E-02 & 6.76E-03 & Tm & 2.06E-06 & 2.40E-06 & 3.40E-07\\
Co & 6.61E-06 & 1.09E-04 & 1.02E-04 & Yb & 7.28E-06 & 2.08E-05 & 1.36E-05\\
Ni & 1.18E-03 & 1.24E-03 & 5.74E-05 & Lu & 1.38E-06 & 8.96E-07 & -4.86E-07\\
Cu & 9.9960E+01 & 9.9173E+01 & -7.8700E-01 & Hf & 8.36E-06 & 1.14E-05 & 3.08E-06\\
Zn & 5.27E-05 & 1.49E-04 & 9.63E-05 & Ta & 3.65E-04 & 4.96E-04 & 1.32E-04\\
Ga & 8.63E-06 & 1.59E-04 & 1.50E-04 & W & 7.59E-06 & 3.41E-03 & 3.40E-03\\
Ge & 1.21E-04 & 5.63E-04 & 4.42E-04 & Re & 1.54E-05 & 7.93E-06 & -7.49E-06\\
As & 2.13E-04 & 8.46E-05 & -1.29E-04 & Os & 4.04E-06 & 6.03E-06 & 1.99E-06\\
Se & 1.19E-04 & 7.37E-04 & 6.18E-04 & Ir & 3.27E-06 & 1.71E-06 & -1.56E-06\\
Br & 2.13E-05 & 3.39E-05 & 1.26E-05 & Pt & 7.58E-06 & 3.18E-06 & -4.40E-06\\
Kr & 1.49E-05 & 1.86E-05 & 3.64E-06 & Au & 3.35E-06 & 1.56E-05 & 1.22E-05\\
Rb & 6.26E-04 & 2.76E-04 & -3.50E-04 & Hg & 4.16E-04 & 9.02E-04 & 4.87E-04\\
Sr & 2.20E-06 & 3.30E-06 & 1.10E-06 & Tl & 5.09E-06 & 1.57E-05 & 1.06E-05\\
Y & 2.65E-06 & 5.18E-06 & 2.53E-06 & Pb & 8.20E-04 & 6.86E-01 & 6.86E-01\\
Zr & 6.11E-06 & 1.20E-05 & 5.90E-06 & Bi & 2.07E-06 & 4.84E-05 & 4.63E-05\\
Nb & 4.33E-06 & 1.81E-05 & 1.38E-05 & Th & 9.93E-07 & 2.84E-06 & 1.85E-06\\
Mo & 6.43E-05 & 6.38E-05 & -4.77E-07 & U & 1.34E-06 & 3.82E-06 & 2.49E-06\\
Ru & 7.17E-06 & 3.98E-06 & -3.19E-06 &                                   \\
\hline
   &          &          &           & \textbf{TOTAL} & 1.00000E+02 & 1.00000E+02 & 7.870E-01\\
\hline \hline
\end{tabular}
\end{table*}

  Prior to the analysis of the derived results, we consider still
one aspect. Under the regimes used in the glow-discharge cell (Ar
plasma) of a mass spectrometer VG~9000, the etching rate of a
specimen was about $0.5\,\mu$m/min. To avoid any effect of surface
contaminations on the results of analysis, we etched the specimen
surface usually for 30\dots 40\,min prior to the registration of a
spectrum every time. The application of this procedure led to the
removal of the surface layer of at least $10\,\mu$m in thickness
which contained, as usual, the enhanced amount of the admixture.

First of all, we will discuss the data given in column
``Increment'' in Table 4. By definition, they are the difference
of the data given in columns ``Sandwich'' and ``Initial.'' It is
clear that these data must have positive sign for all the chemical
elements except for Cu (it became more diluted), because there are
no reasons for that the content of minor chemical elements in the
accumulating screen decreases after the explosion of a target. In
our opinion, there exists only a sole explanation for the
indicated fact of a decrease in concentrations. We note that sheet
materials are, as usual, inhomogeneous by
composition~\cite{Gulya77,Metal87,Mitin87}. In the case of
``sandwich,'' we meet a good averaging of the composition of the
initial material (20 blanks cut from various parts of a sheet).
But, in the case of the initial specimen, the averaging is not
sufficient, because we deal with only one blank. Moreover, as
known~\cite{Gulya77,Metal87,Mitin87}, the chemical inhomogeneity
is usually manifested in a greater degree for those chemical
elements, whose content in the material is low, which happens in
the situation under consideration.

  It is obvious that the sum of all values in column ``Increment''
(without the increment for Cu)
\begin{equation}
\Delta C_{m.a.}=0.787\,\mbox{mass \%}%
\label{Eq20}
\end{equation}
has sense of a change in the concentration of minor chemical
elements in the analyzed region of the accumulating screen due to
the explosion of a Pb target. The increment of the concentration
of Pb (see Table~\ref{Table4})
\begin{equation}
\Delta C_\mathrm{Pb}=0.686\,\mbox{mass \%}.%
\label{Eq21}
\end{equation}

It is clear that we deal with Pb transferred from the target.
However, the target lead transferred else some amount of
impurities contained in it, for which the obvious relation
\begin{equation}
\Delta C_{\mathrm{Pb.}im.} = \frac{\Delta C_\mathrm{Pb} \cdot C_{im.t.}}{100\,\%} \approx 0.0003\,\mbox{mass \%}.%
\label{Eq22}
\end{equation}
is valid. Now, with regard for Eqs.~\ref{Eq20}--\ref{Eq22}, the
concentration of foreign chemical elements in the analyzed region
of the accumulating screen which have appeared due to the
explosion of a target and did not belong to the target prior to
its explosion satisfies the relation
\begin{equation}
 C_{f.a.}=\Delta C_{m.a.}-\Delta C_\mathrm{Pb}-\Delta C_{\mathrm{Pb.}im.} \approx 0.10\,\mathrm{mass}\,\%.%
\label{Eq23}
\end{equation}

One can easily see that in Eq.~\ref{Eq23} we can safely neglect
the third term, i.e., we should not take the admixture transferred
by the target Pb into account while calculating the concentration
of foreign chemical elements in the analyzed region.

  For a better insight into the situation as for foreign
chemical elements, we pass from the quantity $C_{f.a.}$
characterizing their relative amount (it depends on the
accumulating screen thickness) to absolute values: namely, we will
find the number of their atoms and their total mass. The mass of a
part of the accumulating screen, which is located under the area
of the analyzed region, is (see Fig.~\ref{Fig_est03})
\begin{equation}
m_{a.r.}=\frac{\pi d^2}{4} H \rho_\mathrm{Cu} \approx
87.38\,\mbox{mg}.%
\label{Eq24}
\end{equation}

Now we can easily obtain the total mass of all atoms of foreign
chemical elements being in the volume of the analyzed region of
the screen as
\begin{equation}
m_{f.a.} = m_{a.r.} C_{f.a.} \approx 0.09\,\mbox{mg}%
\label{Eq25}
\end{equation}
as well as their total number
\begin{equation}
N_{f.a.}= \sum \limits_i \frac{m_{f.a.}\Delta C_i}{M_i}N_A \approx 1.66\times10^{18},%
\label{Eq26}
\end{equation}
where $i$ is the summation index passing all the chemical elements
from Table~\ref{Table4} except for Cu and Pb, $\Delta C_i$ is the
mass concentration of the $i$-th foreign chemical element in the
analyzed region (the data from column ``Increment'' in
Table~\ref{Table4}), $M_i$ is the molar mass of the $i$-th foreign
chemical element, and $N_A$ is the Avogadro number. For the number
of nucleons being present in the amount of foreign chemical
elements given by Eq.~\ref{Eq26}, we get~\cite{Nuclid}
\begin{equation}
N_\mathrm{nucl}= 5.49\times10^{19}.%
\label{Eq27}
\end{equation}

  A special attention should be paid to the fact that the values
of $m_{f.a.}$ and $N_{f.a.}$ are, in fact, the result of direct
measurements and are referred exclusively to the analyzed region,
i.e., to the full volume of a part of the accumulating screen
which is located under the analyzed area positioned in the central
part of the screen. In this connection, we are tempted to recount
the indicated values for the whole accumulating screen by
multiplying them by the ratio of the squares of the diameters of
the accumulating screen $D^2$ and the analyzed region $d^2$.
However, it is worth noting that the mentioned procedure is
already based on the assumption as for that the distribution of
products of the explosion over the accumulating screen surface is
close to a homogeneous one. Of course, this assumption should be
verified.

  We will implement such a verification in the following way. First,
we calculate the mass of Pb transferred from the target on the
analyzed region:
\begin{equation}
m_{\mathrm{Pb}a} = m_{a.r.} \Delta C_\mathrm{Pb} \approx 0.60\,\mbox{mg}.%
\label{Eq28}
\end{equation}

Now, by supposing a distribution of Pb to be homogeneous, we
calculate its mass transferred from the target on the whole
accumulating screen:
\begin{equation}
m_{\mathrm{Pb}s} = m_{\mathrm{Pb}a}\frac{D^2}{d^2} \approx 5.40\,\mbox{mg}.%
\label{Eq29}
\end{equation}

We note that the derived amount of Pb must be only a fraction of
the target mass loss (we recall that the mean target mass loss
after its explosion $\Delta m_t$ was equal to 2.60\,mg in the
performed experiment) since it was collected only in the solid
angle, at which the accumulating screen is seen from the collapse
center. Moreover, the derived amount of Pb is even a fraction of
the substance collected in the indicated solid angle, because the
screen was covered, besides Pb, by the atoms of foreign chemical
elements (the transformed substance of the target). Hence, by
virtue of the above-presented reasoning, the derived relation
\begin{equation}
m_{\mathrm{Pb}s} > \Delta m_t%
\label{Eq30}
\end{equation}
is inept. Contradiction (Eq.~\ref{Eq30}) obviously indicates that
the assumption used in its derivation as for the homogeneity of
the distribution of products of the explosion over the
accumulating screen surface is erroneous.

\begin{figure}
\centering
\includegraphics[width=8 cm]{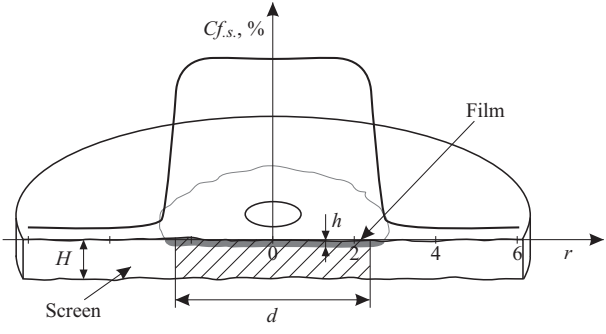}
\caption{Schematic image of the distribution of foreign chemical
elements over the accumulating screen surface.} \label{Fig_est06}
\end{figure}

  Thus, the performed measurements and estimations yield that products
of the explosion of a target are very irregularly distributed over
the accumulating screen surface. Most products find themselves in
the central part of the screen, and their amount sharply drops on
its periphery. In other words, the distribution of foreign
chemical elements over the accumulating screen surface has the
form schematically shown in Fig.~\ref{Fig_est06}.

  There are other facts which not only confirm the above-discussed
distribution of the target substance over the accumulating screen
surface, but correct it. In particular, the composition of the
accumulating screen was determined on the ``sandwich'' not only in
the central part of the analyzed region (see
Fig.~\ref{Fig_est04}), but also on its edge, which corresponds to
the peripheral sections of the accumulating screen. For the total
concentration of the atoms of all foreign chemical elements, here
we get
\begin{equation}
C^{a.}_{f.a.} \approx 0.01\,\mbox{mass \%},%
\label{Eq31}
\end{equation}
i.e., this value is 10 times less than the similar value for the
analyzed region positioned near the center of the screen (see
Eq.~\ref{Eq23}). Basing on Eq.~\ref{Eq31}, it is easy to obtain
the total mass of all the atoms of foreign chemical elements on
the whole accumulating screen as
\begin{eqnarray}
m_{f.s.} &\approx& m_{a.r.} C_{f.a.} +
(\frac{D^2}{d^2}-1)m_{a.r.}C^{a.}_{f.a.} \nonumber\\
&=&1.8 m_{a.r.} C_{f.a.} = 1.8 m_{f.a.} \approx 0.16\,\mathrm{mg}%
\label{Eq32}
\end{eqnarray}

It is obvious that the number of all the atoms of foreign chemical
elements on the whole accumulating screen is more by a factor of
1.8 than their number in the analyzed region near the screen
center, i.e.,
\begin{equation}
N_{f.s.} \approx 1.8N_{f.a.} \approx 3.0\times10^{18}.%
\label{Eq33}
\end{equation}

  The results of measurements of the composition at the ``sandwich''
edge are useful also in another aspect. If we compare them with
those for the ``sandwich'' center, it is noticeable that the ratio
of the concentration of the $i$-th chemical element at the screen
center to that on the screen periphery is approximately equal to
7--15 most chemical elements. This ratio tends rather surely to 10
for those chemical elements, whose concentrations are large, i.e.,
for those which are measured most exactly. We recall that a ratio
of 10 characterizes the total concentrations of the atoms of all
foreign chemical elements at the screen center (Eq.~\ref{Eq23})
and at its edge (Eq.~\ref{Eq31}). In other words, the
above-presented facts indicate that
\begin{equation}
\frac{\Delta C^{c.}_i}{\Delta C^{e.}_i} \approx \frac{C_{f.a.}}{C^{e.}_{f.a.}} \approx 10.%
\label{Eq34}
\end{equation}

Relation (Eq.~\ref{Eq34}) means obviously that the separation of
chemical elements is absent upon the dispersion of a target
substance. In other words, the masses of bits of the substance
removed from the target due to the explosion in different
directions will be different, but the fraction of the mass of any
chemical element contained in them will be identical for all the
directions.

  The absence of the separation of chemical elements upon the dispersion
of a target substance yields that the mass of all the atoms of
foreign chemical elements contained in the substance removed from
the target, $m_f$, is to the mass of all the atoms of foreign
chemical elements come to the analyzed region, $m_{f.a.}$, as the
target mass loss $\Delta m_t$ is to the mass of all the substance
transferred from the target to the analyzed region. Hence, we can
easily get
\begin{equation}
m_f = \frac{\Delta m_t}{m_{f.a.}+m_{\mathrm{Pb}a}}m_{f.a.} \approx 0.34\,\mbox{mg}.%
\label{Eq35}
\end{equation}

By virtue of the same reasoning, we get the similar estimation for
the total number of the atoms of foreign chemical elements
contained in the mass removed from the target:
\begin{equation}
N_f = \frac{m_f}{m_{f.a.}}N_{f.a.} \approx 6.27\times10^{18}.%
\label{Eq36}
\end{equation}

We note that Eq.~\ref{Eq35} and Eq.~\ref{Eq36} characterize the
amount of almost all the substance transformed upon the explosion
of a target. It does not include only the part which was not
removed from the target and remained on the surface of its crater
(see Fig.~\ref{Fig_est01}, a).

  Finally, the absence of the separation of chemical elements upon
the dispersion of the target substance allows us to get one more
important characteristic: the fraction of the transformed
substance in the total amount of the target substance participated
in the explosion, or the efficiency of the process in terms of
mass. It is obvious that this quantity is
\begin{equation}
\eta_\mathrm{Pb}=\frac{m_f}{\Delta m_t}100\% \approx 13\%.%
\label{Eq37}
\end{equation}

  Now we summarize the results of the second investigation with
Pb targets. In this experiment, we analyzed the composition of the
whole accumulating screen prior to the explosion and after it
rather than the composition of the surface layer of the
accumulating screen. The amount of appeared foreign chemical
elements was determined as the difference of the amounts of minor
chemical elements in the two indicated measurements. The used
procedure was, in essence, the direct measurement of the required
value and was carried out so that it excluded absolutely the
interpretation of the fact of enrichment of of the accumulating
screen surface at the expense of the redistribution of impurities
from its bulk. The results of this experiment also exclude the
possibility of generation of foreign atoms on the accumulating
screen due to the transfer of impurities by the target material,
because the first value exceeds the second one by at least 3
orders of magnitude (see Eq.~\ref{Eq22} and Eq.~\ref{Eq23}). It is
easy to understand that the derived amount of foreign atoms cannot
be condensed from the residual atmosphere of the vacuum chamber of
the setup. Indeed, it contained at most $1\times 10^{-5}$\,mg of a
substance (the working liquid vapour, rarefied air, hydrocarbons,
etc.) in a volume of 0.7\,dm$^3$ and at a residual pressure of
about $10^{-3}$\,Pa. Such an amount is again by 4 orders less than
the registered value (see Eq.~\ref{Eq25}).

  Below, we give the estimation of the amount of impurities adsorbed
from air by the surface of an accumulating screen. As known, the
exposure of the extrapure surfaces of solids in air is accompanied
by the formation of a layer of atoms of the chemical elements
contained in air on these surfaces (first of all, they are carbon,
oxygen, and, in the slight amount, nitrogen). Based on the data of
AES~\cite{Briggs87,Czand75,Carl75}, we can infer that their
thickness does not certainly exceed two monolayers of atoms,
because otherwise they would shield almost completely
Auger-electrons going from the substance of the very accumulating
screen (the substrate), which does not occur in reality. It is
easy to calculate that, in the indicated approximation, the amount
of impurities adsorbed from air on the both sides of the
accumulating screen is about $5 \times 10^{-5}$\,mg or $0.6 \times
10^{-4}$~mass~\% in the limits of the analyzed region. The given
estimation yields that neither the composition nor the amount of
air-based impurities allow one to relate the origin of the
discovered atoms of foreign chemical elements on the accumulating
screen to the process of adsorption from air. We also note that
the essentially inhomogeneous distribution of foreign atoms over
the accumulating screen surface does not allow us to explain their
appearance by the process of adsorption from the residual
atmosphere of the vacuum chamber and/or air or by the initial
contamination of the accumulating screen surface. Indeed, in these
cases, we must observe the formation of a more or less homogeneous
distribution of impurities on the screen surface.

  Finally, any transfer of a substance from the shell walls of
the experimental chamber onto an accumulating screen was absent,
because we took care for the suppression of a ricochet of the
dispersed substance of a target from the shell walls onto the
screen. The efficiency of the suppression of this process is
testified by both the directedness of splashes on accumulating
screens (see Fig.~\ref{Fig_est01}, b) and the absence of a
considerable correlation between the composition of the explosion
products on an accumulating screen and the composition of the
material of the shell walls. In other words, the results of the
performed experiment indicate that the atoms of foreign chemical
elements contained in products of the explosion of a target are
nothing else but the target substance undergone a nuclear
transformation. As for the amount of the atoms of foreign chemical
elements on the accumulating screen which was registered in this
experiment (see Eq.~\ref{Eq33}), we note that it agrees well with
the results of the first study (see Eq.~\ref{Eq06}).

  Discussing the last experiment, we should like to mention its
two basic poor aspects. First of all, upon the measurement of the
composition of the initial material of an accumulating screen, its
averaging was carried out only in the limits of one sheet blank.
It is obvious that the quality of this procedure can be
significantly improved, and one can avoid the error defined by the
inhomogeneity of a composition of sheet materials. The second poor
aspect of the last experiment was a low value of the geometric
factor of the used ``sandwich'' equal to 0.004 (see
Eq.~\ref{Eq12}), which forces us to register a rather small
quantity, namely a weak signal from the layer of about $2\,\mu$m
in thickness formed by products of the explosion against the
background of a strong signal from the accumulating screen of
$500\,\mu$m in thickness playing the role of a substrate. Of
course, we can enhance the accuracy of measurements in this
situation.

  In the third experiment, we undertook some measures in order
to eliminate all drawbacks of the two previous ones. As in the
second experiment, we used the highly sensitive method of GDMS
(VG~9000, VG~Elemental, UK). We took ``sandwiches'' as the initial
specimen and the specimen ``processed'' by an explosion, which
ensured the efficient averaging of a composition of the initial
material of accumulating screens. Its geometric factor was
increased up to 0.02, i.e., by 5 times. We reached this value by
decreasing the screen thickness $H$ to $200\,\mu$m (for lower
thicknesses, the screen is broken by the explosion of a target)
and applying the products of explosions on both sides of the
screen with $h \approx 2\,\mu$m. Thus, the region analyzed by a
mass spectrometer on the ``sandwich'' included 25 ends of
accumulating screens for both the initial and ``processed''
specimens. In experiments, we used the target and the accumulating
screen made of Ag and Cu, respectively. By weighting, we found
that the mean mass losses of the target $\Delta m_t$ and the
accumulating screen $\Delta m$ were, respectively, 3.10 and
3.28\,mg (see Table~\ref{Table5}).

  First, we analyzed the initial material of the Ag target by GDMS.
According to the averaged result of two measurements, the total
content of impurities in it was about 0.269 mass \% with Cu as the
main component (0.217 mass \%). Because this copper does not
``contaminate'' the copper accumulating screen, we may estimate
the target admixture transferred onto the accumulating screen
during the explosion, $C_{im.t.}$, as 0.052 mass \% of the amount
of the transferred Ag (the difference of the values given above).

\begin{table}
\centering%
\caption{\label{Table5}Data on mass losses in the experiments with
Ag targets and Cu screens.}
\begin{tabular}{cccccc}
\hline \hline%
Specimen &\multicolumn{2}{c}{Mass loss, mg}&Specimen &\multicolumn{2}{c}{Mass loss, mg} \\
\cline{2-3} \cline{5-6}
 number  &Screen  &Target & number  &Screen  &Target \\
\hline%
8689 & 6.05 & 3.88 & 8663 & 3.14 & 3.48\\
8693 & 3.06 & 4.52 & 8680 & 6.99 & 2.62\\
8621 & 5.18 & 4.15 & 8675 & 1.23 & 1.13\\
8668 & 3.84 & 3.46 & 8652 & 8.37 & 2.16\\
8666 & 0.39 & 2.03 & 8698 & 2.01 & 1.68\\
8683 & 0.60 & 3.29 & 8695 & 2.40 & 4.34\\
8665 & 5.36 & 4.29 & 8651 & 4.24 & 4.09\\
8682 & 4.14 & 4.61 & 8672 & 0.01 & 1.35\\
8653 & 2.57 & 2.26 & 8676 & 3.79 & 4.06\\
8671 & 3.83 & 3.44 & 8654 & 1.12 & 1.01\\
8657 & 0.56 & 2.36 & 8697 & 0.88 & 1.77\\
8679 & 2.14 & 3.33 & 8699 & 2.57 & 2.47\\
8655 & 5.13 & 3.37 & 8656 & 4.68 & 3.56\\
8677 & 2.61 & 2.93 & 8678 & 2.21 & 1.91\\
8688 & 2.76 & 2.46 & 8681 & 1.56 & 1.27\\
8691 & 4.72 & 3.49 & 8664 & 3.44 & 3.08\\
8685 & 2.92 & 3.19 & 8696 & 5.99 & 4.12\\
8687 & 4.67 & 3.80 & 8652 & 2.92 & 6.76\\
\hline%
 &  &  & Average & 3.28 & 3.10\\
\hline \hline
\end{tabular}
\end{table}

In Table~\ref{Table6}, we present the data on changes of the
element composition of accumulating Cu screens after the
experiments with a silver target. All the designations and data
have the same sense as in Table~\ref{Table4}. Each value of the
compositions of the initial and processed specimens is averaged
over 4 measurements. Columns ``Increment'' include the amounts of
minor chemical elements (for each element separately) appeared as
a result of the explosion of a target. We note that, in this case,
there are no negative values. This fact testifies to the
correctness of our assumption as for both the origin of negative
values of the increments for a number of chemical elements in the
previous experiment and the efficiency of solving the problem of
averaging of a chemical inhomogeneity of sheet materials upon the
determination of their composition by means of the use of
specimens of the ``sandwich'' type. The total amount of minor
chemical elements appeared after the explosion is
\begin{equation}
\Delta C_{m.a.} = 5.96\,\mbox{mass \%},%
\label{Eq38}
\end{equation}

This value includes the silver transferred from the target,
\begin{equation}
\Delta C_\mathrm{Ag} = 5.36\,\mbox{mass \%},%
\label{Eq39}
\end{equation}

After the subtraction of it, we get that the amount of a substance
formed as the result of the nucleosynthesis:
\begin{equation}
C_{f.a.} = \Delta C_{m.a.} - \Delta C_\mathrm{Ag} - \Delta C_{\mathrm{Ag.}im.} \approx 0.60\,\mbox{mass \%},%
\label{Eq40}
\end{equation}

In the last value, the admixture transferred by Ag (5.36 mass
\%) from the target is
\begin{equation}
\Delta C_{\mathrm{Ag.}im.} \approx 0.003\,\mbox{mass \%},%
\label{Eq41}
\end{equation}

That is, we can neglect it, because this value does not exceed the
accuracy limits for the measured quantity. It is worth noting that
values Eqs.~\ref{Eq38}--\ref{Eq40} derived by us exceed the values
of the same quantities from the previous experiment (see
Eqs.~\ref{Eq20}, \ref{Eq21}, and \ref{Eq23}) by almost one order.
It is doubtless that this circumstance is caused by the increase
in the geometric factor and favours the increase in the accuracy
of determination of the quantities under study.

\begin{table*}
\centering%
\caption{Changes in the element composition of an accumulating Cu
screen (`\label{Table6}`sandwich'') after the explosion of a
silver target.}
\begin{tabular}{cccccccc}
\hline \hline%
Element &\multicolumn{3}{c}{Concentration, mass \%} &Element& \multicolumn{3}{c}{Concentration, mass \%}       \\
\cline{2-4} \cline{6-8}
        &Initial  &``Sandwich''& Increment& &Initial  &``Sandwich''& Increment \\
\hline%
H & 8.69E-05 & 9.76E-05 & 1.07E-05 & Rh & 2.07E-06 & 8.48E-06 & 6.42E-06\\
Li & 1.28E-04 & 5.38E-04 & 4.10E-04 & Pd & 1.10E-05 & 1.61E-05 & 5.08E-06\\
Be & 3.72E-07 & 6.30E-06 & 5.93E-06 & Ag & 1.57E-03 & 5.37E+00 & 5.36E+00\\
B & 1.79E-04 & 5.84E-04 & 4.05E-04 & Cd & 1.02E-05 & 4.82E-05 & 3.79E-05\\
C & 2.07E-04 & 2.68E-04 & 6.06E-05 & In & 3.02E-06 & 4.85E-06 & 1.84E-06\\
N & 5.53E-04 & 6.52E-04 & 9.91E-05 & Sn & 5.25E-04 & 1.18E-03 & 6.56E-04\\
O & 5.00E-04 & 5.74E-04 & 7.45E-05 & Sb & 4.47E-06 & 6.45E-05 & 6.00E-05\\
F & 4.08E-05 & 4.51E-05 & 4.35E-06 & Te & 3.06E-05 & 7.80E-05 & 4.74E-05\\
Na & 3.85E-03 & 8.50E-03 & 4.66E-03 & I & 8.78E-07 & 1.18E-06 & 2.98E-07\\
Mg & 1.33E-04 & 1.11E-03 & 9.75E-04 & Cs & 4.26E-05 & 1.00E-04 & 5.74E-05\\
Al & 1.04E-03 & 5.26E-02 & 5.15E-02 & Ba & 7.55E-06 & 6.04E-04 & 5.96E-04\\
Si & 3.75E-05 & 8.70E-03 & 8.66E-03 & La & 3.10E-07 & 3.83E-07 & 7.33E-08\\
P & 3.55E-02 & 3.60E-02 & 5.27E-04 & Ce & 4.92E-07 & 5.03E-07 & 1.09E-08\\
S & 6.01E-03 & 1.00E-02 & 3.99E-03 & Pr & 3.66E-07 & 3.84E-07 & 1.77E-08\\
Cl & 6.58E-03 & 4.20E-02 & 3.54E-02 & Nd & 4.35E-06 & 4.48E-06 & 1.30E-07\\
K & 2.44E-05 & 5.58E-05 & 3.14E-05 & Eu & 1.73E-07 & 1.11E-06 & 9.37E-07\\
Ca & 2.79E-04 & 4.20E-02 & 4.17E-02 & Sm & 2.77E-06 & 2.89E-06 & 1.27E-07\\
Sc & 1.26E-06 & 6.65E-06 & 5.39E-06 & Gd & 4.83E-07 & 7.96E-07 & 3.13E-07\\
Ti & 1.21E-04 & 7.75E-04 & 6.54E-04 & Tb & 2.32E-07 & 2.63E-07 & 3.14E-08\\
V & 2.15E-05 & 1.78E-04 & 1.57E-04 & Dy & 1.66E-06 & 1.74E-06 & 7.81E-08\\
Cr & 3.16E-05 & 4.02E-04 & 3.71E-04 & Ho & 8.13E-08 & 4.55E-07 & 3.74E-07\\
Mn & 8.34E-05 & 8.64E-05 & 2.98E-06 & Er & 2.84E-06 & 8.95E-06 & 6.11E-06\\
Fe & 6.70E-03 & 1.25E-01 & 1.18E-01 & Tm & 2.61E-07 & 3.40E-07 & 7.85E-08\\
Co & 5.23E-05 & 1.44E-04 & 9.17E-05 & Yb & 1.83E-06 & 2.40E-06 & 5.73E-07\\
Ni & 1.34E-03 & 2.42E-03 & 1.08E-03 & Lu & 3.47E-07 & 3.78E-07 & 3.14E-08\\
Cu & 9.9631E+01 & 9.3706E+01 & -5.9250E+00 & Hf & 7.29E-07 & 6.79E-06 & 6.06E-06\\
Zn & 2.96E-01 & 5.52E-01 & 2.56E-01 & Ta & 1.59E-06 & 1.79E-03 & 1.79E-03\\
Ga & 3.28E-05 & 3.91E-05 & 6.37E-06 & W & 5.06E-03 & 3.86E-02 & 3.36E-02\\
Ge & 1.62E-05 & 3.19E-05 & 1.57E-05 & Re & 1.83E-07 & 1.86E-06 & 1.68E-06\\
As & 2.11E-06 & 3.00E-06 & 8.89E-07 & Os & 1.01E-06 & 2.52E-05 & 2.42E-05\\
Se & 1.44E-04 & 2.00E-04 & 5.63E-05 & Ir & 6.63E-07 & 1.07E-06 & 4.04E-07\\
Br & 8.03E-06 & 1.25E-05 & 4.47E-06 & Pt & 8.74E-07 & 9.05E-07 & 3.15E-08\\
Rb & 2.48E-06 & 2.49E-06 & 1.25E-08 & Au & 1.48E-06 & 3.32E-06 & 1.85E-06\\
Sr & 2.34E-07 & 1.19E-06 & 9.58E-07 & Hg & 4.23E-05 & 5.31E-03 & 5.27E-03\\
Y & 5.73E-07 & 5.76E-07 & 3.60E-09 & Tl & 4.90E-07 & 7.99E-06 & 7.50E-06\\
Zr & 1.98E-06 & 3.19E-06 & 1.21E-06 & Pb & 7.10E-04 & 3.40E-02 & 3.33E-02\\
Nb & 6.37E-06 & 8.00E-06 & 1.63E-06 & Bi & 2.60E-05 & 2.97E-04 & 2.71E-04\\
Mo & 1.52E-05 & 4.06E-04 & 3.91E-04 & Th & 2.49E-07 & 9.85E-07 & 7.36E-07\\
Ru & 1.09E-06 & 1.87E-06 & 7.81E-07 & U & 1.93E-07 & 3.37E-07 & 1.43E-07\\
\hline
 &  &  &  & \textbf{TOTAL} & 1.000E+02 & 1.000E+02 & 5.96E+00\\
\hline \hline
\end{tabular}
\end{table*}

  By using the procedure described above, we calculated the total
mass of all the atoms of foreign chemical elements formed as the
result of the nucleosynthesis and presented in the volume of the
analyzed region of the screen,
\begin{equation}
m^\ast_{f.a.} = m_{a.r.}C_{f.a.} \approx 0.21\,\mbox{mg}.%
\label{Eq42}
\end{equation}
and their total number
\begin{equation}
N^\ast_{f.a.} \approx 2.36 \times 10^{18}.%
\label{Eq43}
\end{equation}

We note that this number of atoms was synthesized in two series of
experiments since the nucleosynthesis products were applied on
both sides of each accumulating screen. Hence, in the performed
experiments, we registered
\begin{equation}
N_{f.a.} = \frac{N^\ast_{f.a.}}{2} \approx 1.18\times10^{18}%
\label{Eq44}
\end{equation}
synthesized atoms on an accumulating screen in the area of 5\,mm
in diameter which is positioned in the central region. Their mass
\begin{equation}
m_{f.a.}=\frac{m^\ast_{f.a.}}{2} \approx 0.105\,\mbox{mg}.%
\label{Eq45}
\end{equation}

As for nucleons contained in the amount of foreign atoms equal to
Eq.~\ref{Eq44}, their number
\begin{equation}
N_\mathrm{nucl}=5.78\times10^{19}.%
\label{Eq46}
\end{equation}

The relative error of the measured values is in the limits of
10--20\%. Comparing Eq.~\ref{Eq45} and Eq.~\ref{Eq25}, we note
that the mass of regenerated atoms per explosion is almost the
same for Pb and Ag targets, but their number is considerably less
for Ag targets (see Eq.~\ref{Eq44} and Eq.~\ref{Eq26}). This fact
indicates that the explosion of an Ag target generates, on the
average, the atoms of heavier chemical elements.

 We also present the mass of Ag transferred from the target onto
the analyzed region as the result of one explosion:
\begin{equation}
m_{\mathrm{Ag}a} = \frac{m_{a.r.}\Delta C_\mathrm{Ag}}{2} \approx 0.94\,\mathrm{mg}.%
\label{Eq47}
\end{equation}

Unfortunately, we did not perform measurements on the edge of the
``sandwich'' in the last experiment, because we did not attach any
special significance to them at that time. The results of
measurements gave small numbers of the atoms of foreign chemical
elements and seemed to be unconvincing for the justification of
the running of the nucleosynthesis reactions upon the explosion of
a target. Therefore, basing on these results, we cannot calculate
and present the values characterizing the amount of the
regenerated substance deposited on the whole accumulating screen.
However, in view of the absence of separation of chemical elements
upon the dispersion of a target substance, we estimated the mass
of all the synthesized atoms contained in the mass removed from
the target
\begin{equation}
m_f \approx 0.31\,\mathrm{mg}.%
\label{Eq48}
\end{equation}

Their number
\begin{equation}
N_f \approx 3.48\times 10^{18}.%
\label{Eq49}
\end{equation}

  Finally, for this experiment, the parameter of efficiency of
the process of transformation of the target substance in terms
of mass is
\begin{equation}
\eta_\mathrm{Ag} \approx 10\,\%.%
\label{Eq50}
\end{equation}

Comparing values Eqs.~\ref{Eq50} and \ref{Eq37}, we emphasize the
lower efficiency of the process of transformation for a silver
target.

  By summarizing, we note that three independent experiments which
were performed by two methods (the first method is local, and the
second is highly sensitive integral) established the fact of the
appearance of foreign chemical elements in products of the
explosion of a solid target. We have demonstrated that these atoms
are the products of a nuclear transformation of the target
substance subjected to the action of a superintense impact
compression. The estimates of the amounts of the substance formed
due to the nuclear transformation well agree with one another.

\section{CONCLUSIONS}

  The main results and conclusions concerning the experimental
investigations of the explosion-induced destruction of solid
targets subjected to the compression up to superhigh densities,
which were performed by XEPMA and GDMS in the present work are as
follows.

\begin{itemize}
\item By the example of three independent analyzes performed by
XEPMA and GDMS, we have demonstrated the fact of discovery of the
atoms of foreign chemical elements in products of the explosion of
a solid target. Prior to the explosion, these elements did not
belong to the materials of the target and screen (the structural
details of the experimental chamber which participated in the
process of explosion). We have shown that their origination is not
related to the processes of deposition from the residual
atmosphere of the vacuum chamber or to the transfer from the shell
walls (the structural details which were present in the
experimental chamber but did not participate in the process of
explosion) but is the result of a nuclear transformation of the
target substance due to the impact compression up to superhigh
densities.%
\item By direct measurements, we determined the amounts of the
nucleosynthesis products deposited on accumulating screens in the
experiments with copper, lead, and silver targets. Their total
mass and the total number of their atoms are, respectively,
$0.1\dots0.2$\,mg and $(1.2\dots3.0) \times 10^{18}$. The
estimates of the amount of a substance formed as the result of a
nuclear transformation which were derived by different methods
are in good agreement with one another.%
\item We have investigated the distribution of the nucleosynthesis
products deposited on the accumulating screen and shown that they
are present as microparticles and films in the surface layer of
about $2\,\mu$m in thickness. Moreover, their overwhelming part is
positioned in the central region of the screen, about 5\,mm in
diameter. Their concentration drops significantly towards the
screen periphery.%
\item We have shown that no separation of chemical elements occurs
in the explosion products deposited on
accumulating screens upon the dispersion of a target substance.%
\item We have determined the quantitative characteristics of the
efficiency for the process of nuclear transformation of a target
substance. For example, the fractions of the regenerated mass of a
target participated in the explosion are 13 and 10\,\% for the Pb
and Ag targets, respectively.
\end{itemize}

\end{document}